\documentclass[preprint]{vgtc}               

\usepackage{mathptmx}
\usepackage{graphicx}
\usepackage{times}

\usepackage{balance}  
\usepackage[T1]{fontenc}
\usepackage{txfonts}
\usepackage{mathptmx}
\usepackage[pdftex]{hyperref}
\usepackage{color}
\usepackage{booktabs}
\usepackage{textcomp}
\usepackage{soul}

\onlineid{0}

\vgtccategory{Research}

\vgtcinsertpkg

\title{Text Entry in Immersive Head-Mounted Display-based Virtual Reality using Standard Keyboards}

\author{Jens Grubert\thanks{e-mail: jg@jensgrubert.de}\\ %
        \scriptsize  Coburg University of Applied Sciences and Arts
\and Lukas Witzani\thanks{e-mail: lukas.witzani@uni-passau.de}\\ %
     \scriptsize University of Passau 
\and Eyal Ofek\thanks{e-mail: eyalofek@microsoft.com}\\ %
     \scriptsize Microsoft Research
\and Michel Pahud\thanks{e-mail: mpahud@microsoft.com}\\ %
     \scriptsize Microsoft Research
\and Matthias Kranz\thanks{e-mail: matthias.kranz@uni-passau.de}\\ %
     \scriptsize University of Passau
\and Per Ola Kristensson\thanks{e-mail: pok21@cam.ac.uk}\\ %
     \scriptsize University of Cambridge}

\teaser{
	\centering
	\\
	\includegraphics[width=0.5\columnwidth]{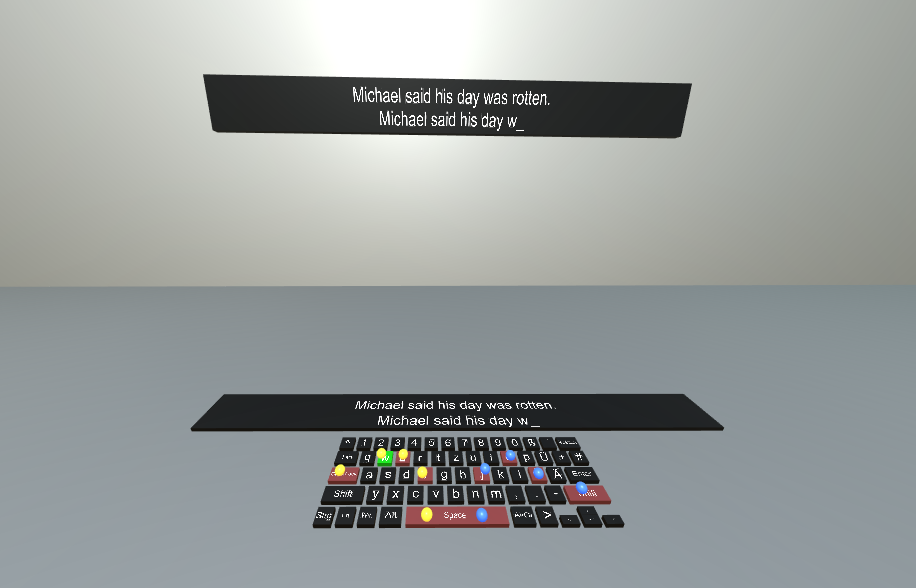}
	\includegraphics[width=0.5\columnwidth]{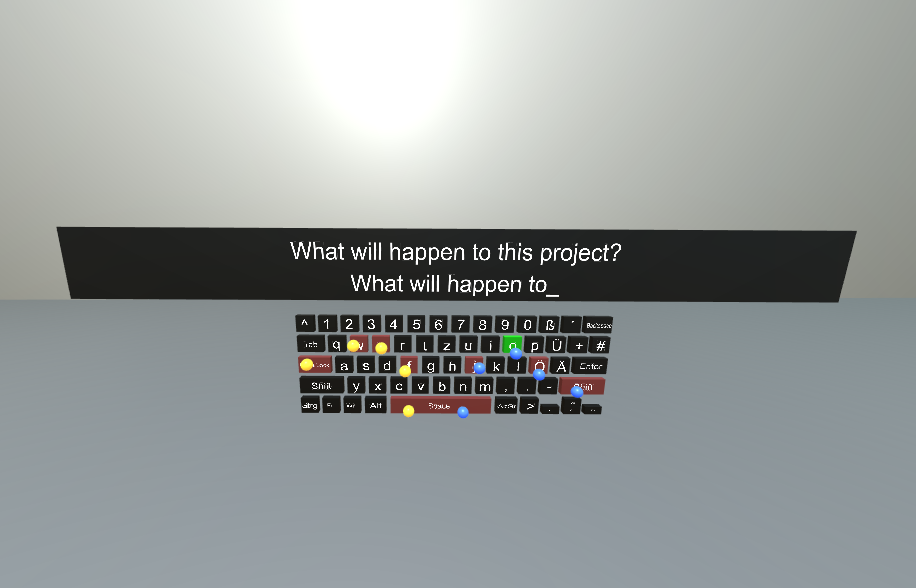} 
	\includegraphics[width=0.5\columnwidth]{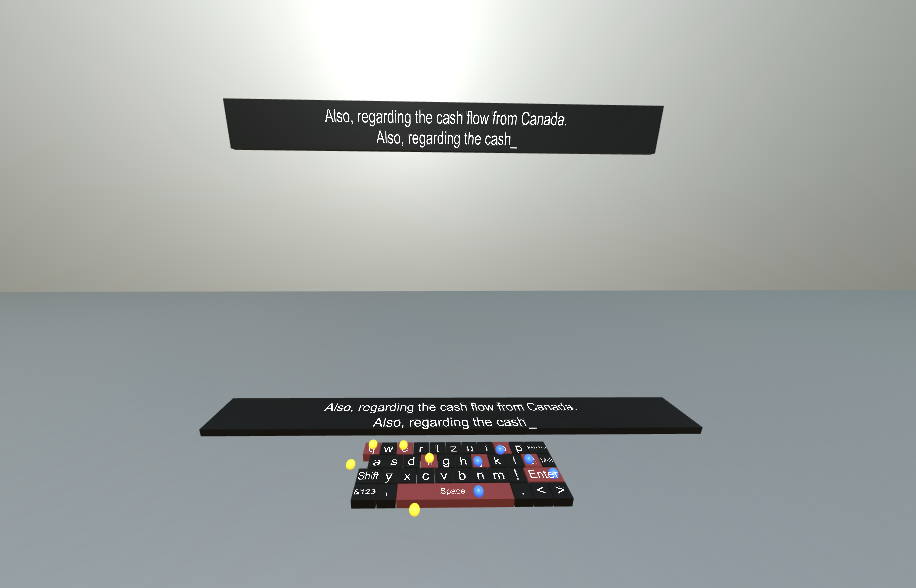}
	\includegraphics[width=0.5\columnwidth]{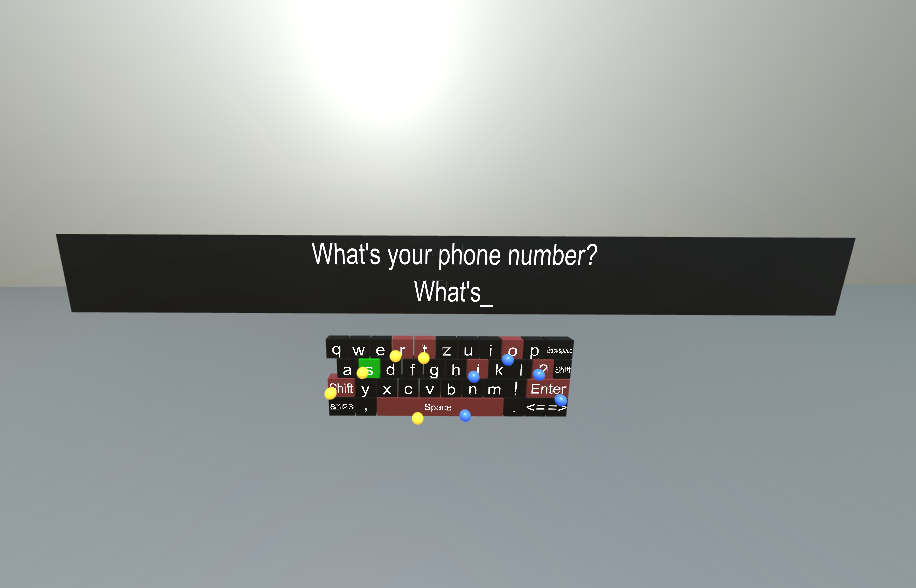} 
	\caption{Conditions studied in the experiment. From left to right: VR views on the conditions \textsc{DesktopKeyboard+NoReposition}, \textsc{DesktopKeyboard+Reposition}, \textsc{TouchscreenKeyboard+NoReposition}, \textsc{TouchscreenKeyboard+Reposition}. 
	}
	\label{fig:conditionsinternal}
}

\abstract{We study the performance and user experience of two popular mainstream text entry devices, desktop keyboards and touchscreen keyboards, for use in Virtual Reality (VR) applications. We discuss the limitations arising from limited visual feedback, and examine the efficiency of different strategies of use. We analyze a total of 24 hours of typing data in VR from 24 participants and find that novice users are able to retain about 60\% of their typing speed on a desktop keyboard and about 40--45\% of their typing speed on a touchscreen keyboard. We also find no significant learning effects, indicating that users can transfer their typing skills fast into VR. Besides investigating baseline performances, we study the position in which keyboards and hands are rendered in space. We find that this does not adversely affect performance for desktop keyboard typing and results in a performance trade-off for touchscreen keyboard typing.
} 

\CCScatlist{ 
  \CCScat{H.5.2:}{ User Interfaces - Input devices and strategies.}{}{}
}

\begin{document}



\maketitle

\section{Introduction}



Alphanumeric text entry is a major interface for many content production applications, from document editing, programming, spread sheet editing, e-mail, web browsing, social communication and many more.
For effective work, people need a keyboard, mostly physical one or a touch screen one, for typing, and a monitor that is big enough to display the edited document well. As the size and resolution of the monitor has a major effect on the effectiveness of the work, it is common to see work stations with one or more large area, high resolution screens. However, there might be different occasions where severe space limitation prevent the use of large screens, such as when traveling on an airplane, or using tiny touchdown work spaces.

Virtual Reality (VR) enables the immersion of the user in graphic content, and may be used to simulate large displays, all around the user, blocking any outside world distraction, making the space appearing larger than it is, and may require a small stand alone headset, ideal for travel. However existing consumer VR systems, such as HTC Vive, Oculus Rift, or Samsung's Gear VR, can only support text entry using hand held controllers, head or gaze direction. Such  methods are tedious and slow, and usually are used to enter very short  texts, such as passwords and names.  Some past works have suggested  dedicated text entry devices, such as wearable gloves \cite{bowman2002text}, or  specialized controllers \cite{Bowman:2004:UIT}, or even a drum set metaphor (a Google Daydream app) which may be found to be efficient, but may require a substantial learning curve and could result in fatigue quickly due to the comparably large spatial movements involved. 


Another challenge, unique for current virtual reality headsets, is the limited angular resolution. The wish to display a large horizontal field of  view, typically of 90 degrees or more diagonally, is addressed by warping a planar screen  display using lenses. The spreading of the screen resolution over such  a large view angle reduce the angular resolution of the display, in  particular around the boundaries of the display. Some Head Mounted  Displays (HMDs) uses Fresnel lens to reduce size and weight of the  headset, which reduces the display quality further more. 
Display of text and the letters over keyboard keys requires a sharp display of  high resolution. Whenever current systems need to display virtual keyboards, they do so over a limited field of view much smaller than the view angle in which a physical keyboard is seen by our eyes in the real world. An attempt to display keyboards in the virtual world at a scale that will fit the user hands should deal with this limitation.

On the other hand, the vertical field of view of common HMDs, is limited; typically to 35\textdegree~downward from the center of the display to the bottom, compared to the large vertical field of view of the human visual system, typically 75\textdegree~downward from the nose.
The natural location of physical text entry devices, lying on a table in front of the user, is not  visible in the HMDs display when the user looks horizontally straight ahead (e.g., in a desktop typing scenario). To be able to see the corresponding virtual representation of these devices, at good resolution, the user has to rotate her head down to face them. This pose can be potentially uncomfortable, as well as stray the user's view from the main, scene which might cause the user to lose the context of the task.

\begin{figure}[!b]
	\centering
	\includegraphics[width=0.8\columnwidth]{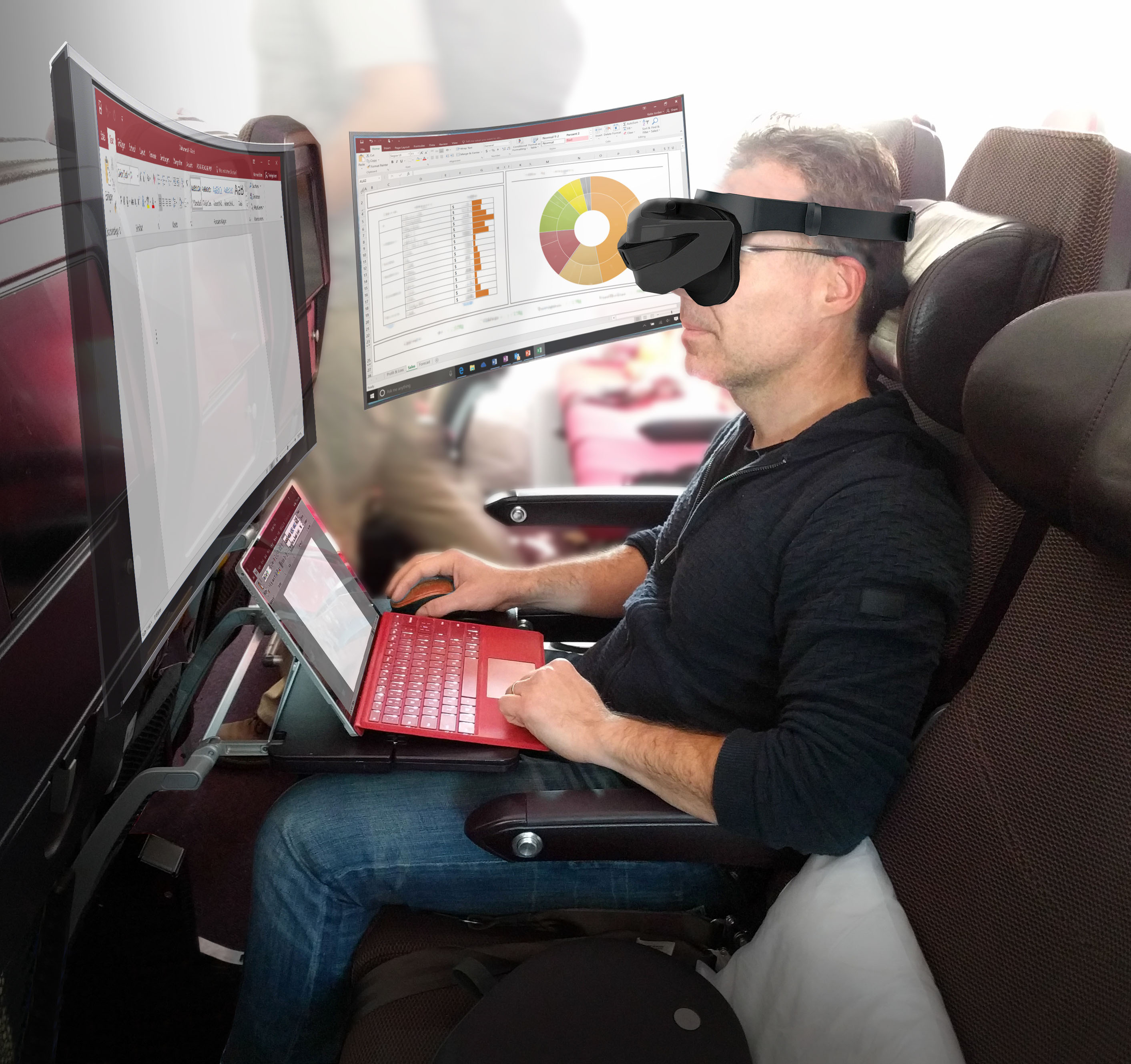}
	\caption{A vision of the future mobile information worker. The user can benefit from typing abilities provided by a laptop or tablet and VR provides an immersive virtual multi-display environment.}
	\label{fig:vision}
\end{figure}

Most current keyboards are not designed for mobility. However, we do see many VR applications where the users sit in a rather static location, in front of a desk. For other applications, small physical or touch keyboards may be attached to the user's non dominant hand. In this paper, we are focusing on extensive text entry capability, so we used the former settings for our research. We can foresee the use of VR by the future information worker, freeing the limitation of physical screens, enabling both 2D and 3D applications and visualizations as a virtual screen environment (see Figure \ref{fig:vision}). However, there is a need for efficient and non-fatiguing text input for use in VR.

Recent research has investigated the feasibility of typing on a physical full-sized keyboard (hereafter referred to as a desktop keyboard) in VR. An obvious problem is the lack of visual feedback. McGill et al.~\cite{mcgill2015dose} found that without visual feedback users' typing performance degraded substantially. However, by blending video of the user's hands into virtual reality the adverse performance differential significantly reduced. An orthogonal approach was explored by Walker et al.~\cite{walker2017efficient}, who explored supporting desktop keyboard typing in VR with a completely visually occluded keyboard. They discovered that their participants typed at an average entry rates of 41.2--43.7 words per minute (wpm), with average character error rates of 8.3\%--11.8\%. These character error rates were reduced to approximately 2.6\%-4.0\% by auto-correcting the typing using the VelociTap decoder \cite{vertanen2015velocitap}.

While transplanting standard keyboards to VR is viable, there are design parameters that are currently still unknown that can plausibly affect performance. First, the performance differential and skill transfer afforded by a desktop keyboard or touchscreen keyboard in VR is not well-understood. While both techniques benefit from being familiar to users, they also provide certain advantages and disadvantages. A desktop keyboard provides tactile sensation feedback of the keys to the user, thereby potentially lessening the importance of visually conveying the location of the user's fingertips in relation to the keyboard in the VR scene. On the other hand, touchscreen keyboard input carried out via a tablet allow for a user interface that can be easily reconfigured and support additional user interface actions, such as crossing, steering and gesturing. Hence, the choice between a desktop and touchscreen keyboard is a cost-benefit decision, which benefits from a clear understanding of the performance implications of a particular choice. 

Second, the keyboard and a visual indication of the user's fingertips can be rendered in different 
locations in VR, which might affect ergonomics and typing speed. In common implementations, 
the virtual representation of the keyboard and hands are aligned with the actual physical input device and the user's hands. This typically induces the need for non-touch typists to look down, as this is where the input device is typically located. However, it is possible to conceive alternative virtual representations, which may encourage a better posture, and allow the user to view both keyboard, hands and entered text, while maintaining eye contact with the VR experience content. It may even allow the incorporation of the keyboard as part of the VR scene, e.g., as an entry control keyboard mounted next to a door. These alternative representations involve transforming the coordinates of the virtual representation of the keyboard and the user's fingertips so that they are no longer aligned to their physical equivalents.

\subsection{Contribution}

In this paper, we present an experiment that investigates desktop and touchscreen keyboard typing performance in VR. Please note that our primary interest is in determining the performance envelope for both keyboard types, as it is known that text entry on standard desktop keyboards is more efficient compared to touch screen keyboards.   We track and render the user's finger tips and a virtual representation of the keyboard. The rendering is minimal to maximize keyboard visibility while still giving a feedback on the fingertip positions. We find that novice users are able to retain about 60\% of their typing speed on a desktop keyboard and about 40--45\% of their typing speed on a multitouch screen virtual keyboard.

In addition, in the same experiment, we examine the effect of relocating the representation of the keyboard and the user's hands in front of the user's view and away from their physical position. We find that this does not adversely affect performance for desktop keyboard typing and results in a performance trade-off for touchscreen keyboard typing.

\begin{figure*}[!t]
	\centering
	\includegraphics[width=0.35\columnwidth]{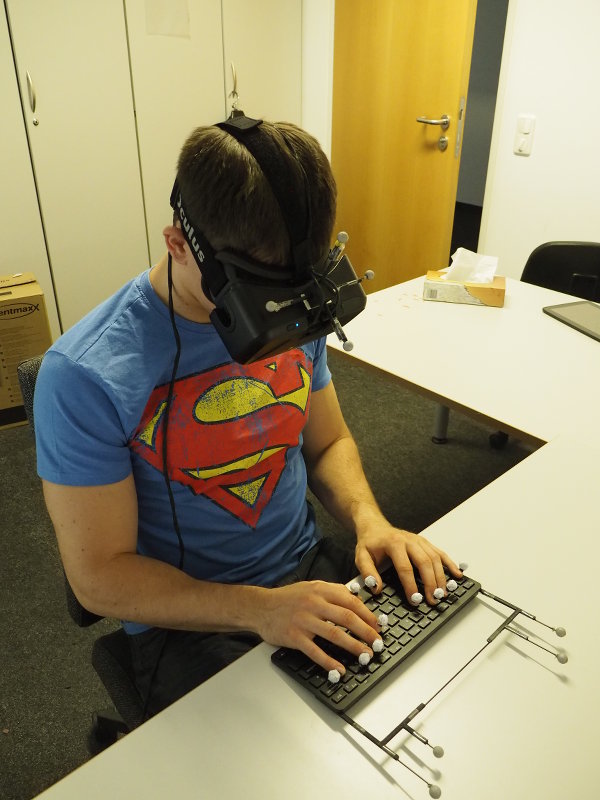}
	\includegraphics[width=0.35\columnwidth]{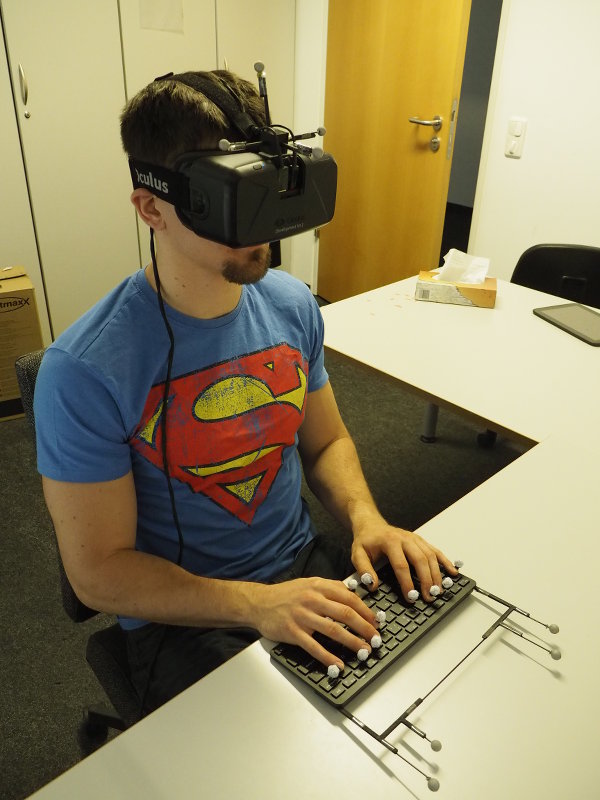} 
	\includegraphics[width=0.35\columnwidth]{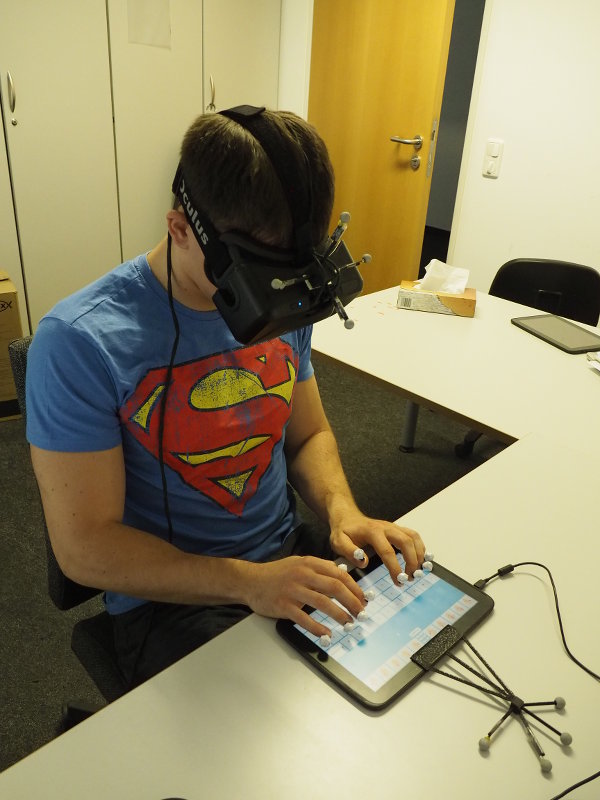}
	\includegraphics[width=0.35\columnwidth]{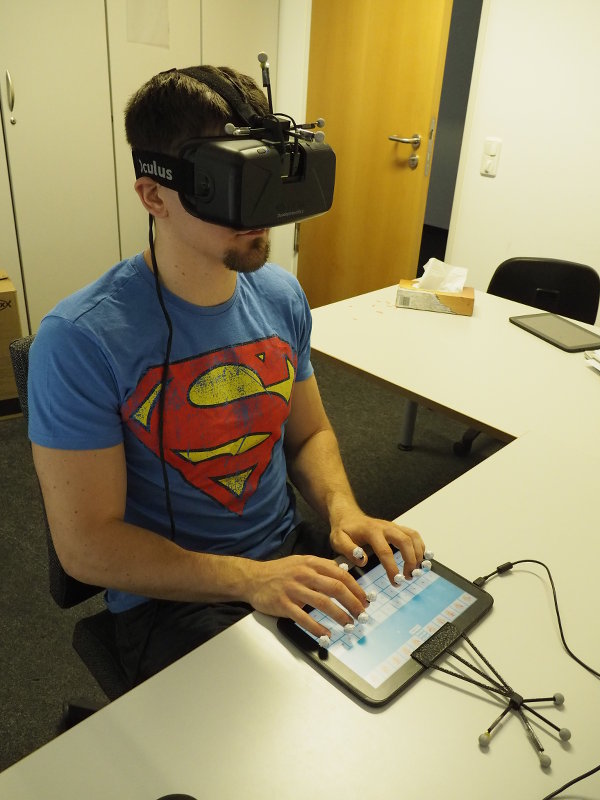}
	\includegraphics[width=0.505\columnwidth]{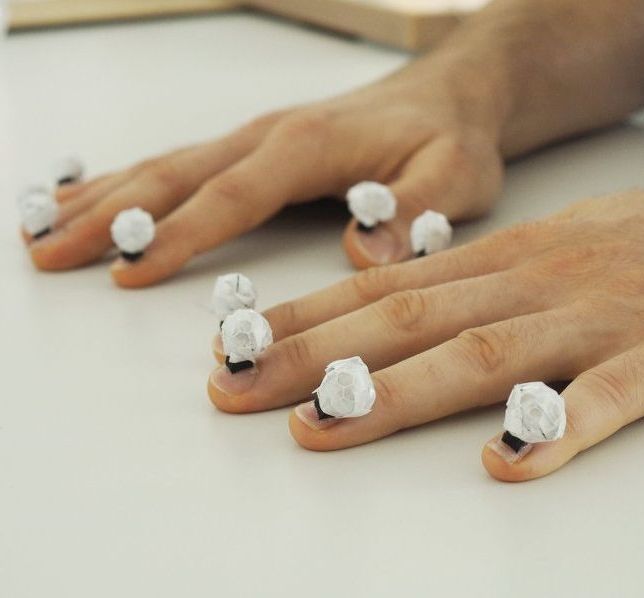}
	\caption{From left to right: external views on the conditions \textsc{DesktopKeyboard+NoReposition}, \textsc{DesktopKeyboard+Reposition}, \textsc{TouchscreenKeyboard+NoReposition}, \textsc{TouchscreenKeyboard+Reposition}, retroreflective markers used for tracking fingers.}
	\label{fig:conditionsexternal}
\end{figure*}



\section{Related Work}



Text entry methods have been widely researched for a variety of application domains, such as mobile text entry 
\cite{mackenzie2002text,zhai2004search} and eye-typing (e.g.,~\cite{majaranta2002twenty}). Often these methods use sophisticated algorithms to infer or predict users' intended text and as a result recent work has focused on designing user interfaces that can support users' uncertain interactions with such interfaces \cite{kristensson2015next,kristensson2009five}.

Relatively few text entry methods have been proposed for VR. Bowman et al.~\cite{bowman2002text} speculate that the reason for this was that symbolic input may seam inappropriate for the immersive VR, and a belief that speech will be the one natural technique for symbolic input. Indeed, when comparing available techniques, they found speech to be the fastest medium for text entry at about 14 words-per-minute (wpm), followed by using a tracked stylus to select characters on a tablet (up to 12 wpm), a specially dedicated glove that could sense a pinch gesture between the thumb and each finger at 6 wpm, and last a commercial chord keyboard which provided 4 wpm. While voice control is becoming a popular input modality \cite{pick2016swifter}, it has severe limitations of ambient noise sensitivity, privacy, and possible obtrusiveness in a shared environment. 

It has also been argued that speech may interfere with the cognitive processes of composing text \cite{shneiderman2000limits}. Also, while dictation of text may feel natural, it is less so for text editing. Correcting speech recognition errors is also a challenge (see Vertanen \cite{vertanen2009efficient} for a recent overview). In addition, speech might not be appropriate in situations such as inside an airplane or when working in cubicles.

Prior work has investigated a variety of wearable gloves, where touching between the hand fingers may represent different characters or words, e.g., \cite{bowman2002text, Hsieh:2016:DWH, Kuester:2005, pratorius2015sensing}. Such devices enable a mobile, eyes-free text entry. However, most of them require a considerable learning effort, and may limit the user ability to use other input devices while interacting in VR, such as game controllers.

Yi et al.~\cite{Yi:2015} suggest a system that senses the motion of hands in the air, to simulate typing. The system PalmType \cite{Wang:2015:PUP} allow a user to use their palm as typing surface.

The standard mobile phone touchscreen keyboard might be a suitable text entry method given that it is portable and provide a relatively high text entry rate with acceptably low error rates \cite{gonzalez2009evaluation, reyal2015performance}. HoVR-type proposes using the hover function of some smartphones as a typing interface in VR \cite{kim2016hovertype} but no text entry rate is reported. 

Head-based text entry in VR also has been investigated. Gugenheimer et al. \cite{gugenheimer2016facetouch} used head-mounted touch screens to enable typing on the face of the user.  Yu et al. \cite{yu2017tap} studied a combination of head-pointing-based text entry with gesture-word recognition. 
Walker et al.~\cite{walker2016decoder} presented the results of a  study of typing on a desktop keyboard with the keyboard either visible or occluded, and while wearing a VR HMD with no keyboard display. They found that the character error rate (CER) was unacceptably high in the HMD condition (7.0\% average CER) but could be reduced to an average 3.5\% CER using an auto-correcting decoder. A year later, they showed that feedback of a  virtual keyboard in VR, showing committed types, can help users correct their hand positions and reduce error rates while typing \cite{walker2017efficient}. In contrast, in this paper, we will show that by visualizing users' finger tips while typing, there is no need for an auto-correcting decoder as with the visual feedback users' character error rate is already sufficiently low for both desktop and touchscreen keyboard typing. 

McGill et al.~\cite{mcgill2015dose} investigated typing on a desktop keyboard in Augmented Virtuality \cite{milgram1994taxonomy}. Specifically, they compared a full keyboard view in reality with a no keyboard condition, a partial and full blending condition. For the blending conditions the authors added a camera view of a partial or full scene into the virtual environment as a billboard without depth cues. They found, that providing a view of the keyboard (partial or full blending) has a positive effect on typing performance. Their implementation is restricted to typing with a monoscopic view of the keyboard and hands and the visualization of hand movements is bound by the update rate of the employed camera (typically 30 Hz). 


In past years, researchers looked at the possibility of decoupling the haptic or tactile sensation feedback from the visual input. By tricking the hand-eye coordination, Azmandian et al.~\cite{Azmandian:2016:HRD} redirected users hands to touch physical objects in a different physical location than the position of their virtual counterparts. We look at the possibilities of displaying hands and keyboards in new virtual positions, different than their physical position, to maximize visibility and utility in VR environment.

Further, Teather et al.~\cite{teather2009evaluating} evaluated the effects of co-location of control and display space in fishtank VR and found only subtle effects of co-location in an object movement task using a tracked stylus. Similarly, further research on visuo-motor co-location on 3D spatial tasks resulted in inconclusive results, not indicating statistically significant differences \cite{fu2012effect, fluet2012effects}.

In this work, we looked at the ability to enter a substantial amount of text in VR while using standard desktop keyboard setups, that are commonly available and require little, if any, learning to use in VR. 



\vspace{-0.15cm}
\section{Repositioning Experiment}

We carried out an experiment to understand the performance of standard keyboards (e.g.~using a keyboard layout the user is familiar with, such as QWERTY, QWERTZ, AZERTY, etc.) in VR. We investigated the performance of desktop keyboard and touchscreen typing. Please note that, while we use a null hypothesis significance testing framework, our primary interest is in determining the performance envelope for both keyboard types. Further, our interest is in studying whether the effect of repositioning the rendering of the text entry device  (e.g.~the desktop keyboard or the touchscreen keyboard) and user's hands representation in VR would affect performance.

\vspace{-0.15cm}
\subsection{Method}
The experiment was a 2 $\times$ 2 within-subjects design with two independent variables with two levels each: \textsc{KeyboardType} and \textsc{VirtualKeyboardPosition}.
The independent variable \textsc{KeyboardType} had two levels: typing on a desktop keyboard (\textsc{DesktopKeyboard}) and typing on a touchscreen keyboard (\textsc{TouchscreenKeyboard}). The independent variable \textsc{VirtualKeyboardPosition} had two levels which affected the position of the keyboard and hands in VR. In the condition \textsc{NoReposition} mode, users' hands and the keyboard virtual representation would appear aligned with the physical (desktop or touchscreen) keyboard and hands. In this condition, the  text was additionally shown in front of the user to not force touch typists to look down to the keyboard. In the \textsc{Reposition} condition, the keyboard and hands would be spatially transformed such that they would be initially visible at the center of the user's field of view, and then fixed in space if users would move their heads. In this condition, the text was shown above the repositioned keyboard. The four conditions are depicted in Figures \ref{fig:conditionsexternal} and \ref{fig:conditionsinternal}. The order of the conditions was counterbalanced. The experiment was carried out in a single 130-minute session structured as a five-minute introduction, 30 minutes of calibration data collection, an 80-minute testing phase (15 minutes per condition + five-minute breaks and questionnaires in between), and 15 minutes for final questionnaires, interviews and debriefing.

\vspace{-0.2cm}
\subsection{Participants}
We recruited 27 participants from a university campus with backgrounds in various fields of study. All participants were familiar with QWERTZ (German keyboard layout) desktop keyboard typing and QWERTZ touchscreen keyboard typing. Two participants had to abort the study, one due to simulator sickness, one due to excessive eyestrain. One participant had to be excluded due to logging issues. From the 24 remaining participants (14 female, 10 male, mean age 23.8 years, sd = 3.1, mean height 174 cm, sd = 11), 15 indicated to have never used a VR HMD before, 6 to have worn a VR HMD once and 3 rarely but more than once. Nine participants indicated to not play video games, 9 rarely, 4 occasionally and 2 very frequently. Ten participants indicated to be highly efficient in typing on a desktop keyboard (6 for virtual keyboard
), 12 to be medium efficient (13 for virtual keyboard) and 2 to write with low efficiency on a desktop keyboard (5 for virtual keyboard). Fifteen participants wore contact lenses or glasses. The volunteers have not participated in other VR typing experiments before.


\subsection{Apparatus and Materials}
\begin{figure}[!b]
	\centering
	\includegraphics[width=0.47\columnwidth]{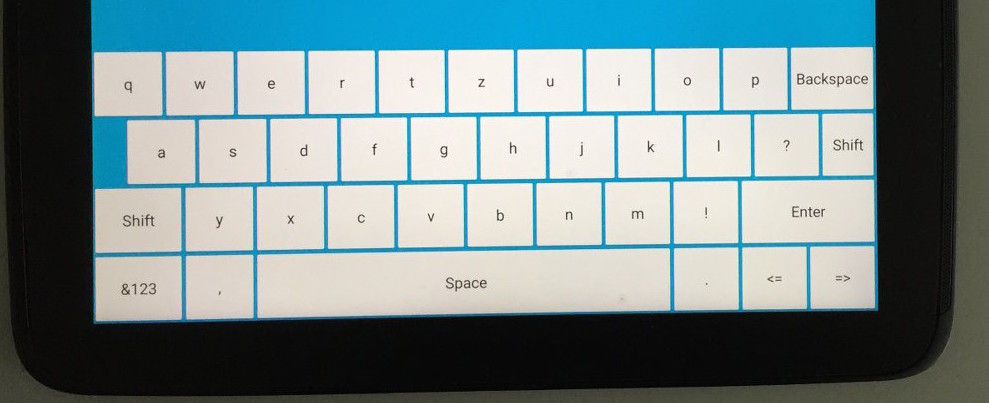} \includegraphics[width=0.45\columnwidth]{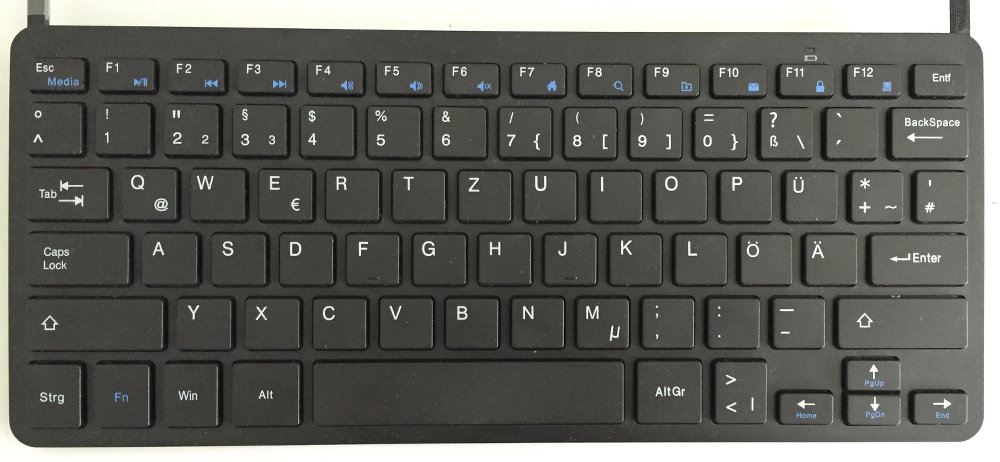}
	\caption{Keyboards used in the experiment. Left: touchscreen keyboard, Right: desktop keyboard.}
	\label{fig:keyboards}
\end{figure}

Stimulus sentences were drawn from the mobile email phrase set \cite{vertanen2011versatile}. Participants were shown stimulus phrases randomly drawn from the set. 
An OptiTrack Flex 13 outside-in tracking system was used for spatial tracking of finger tips and the HMD. It had a mean spatial accuracy of 0.2 mm. An Oculus Rift DK2 was used as HMD. Participants' finger tips were visualized as semi-transparent spheres (c.f. \cite{grubert2018effects}). 
Visual feedback of the left hand were shown in yellow color, whereas the finger tips visual feedback of the right hand were shown in blue color, see Figure \ref{fig:conditionsinternal}. 
In addition, when hovering with fingers above keys, the corresponding key would be highlighted, as seen Figure \ref{fig:conditionsinternal}. The desktop keyboard was a CSL wireless keyboard 
with physical dimensions of (width $\times$ height) 
272 $\times$ 92 mm and key dimensions of 15 $\times$ 14 mm, see Figure \ref{fig:keyboards}, right. The touchscreen keyboard was implemented on a Google Nexus 10 tablet with dimensions of 125 $\times$ 73 mm and key dimensions of 18 $\times$ 17 mm, see Figure \ref{fig:keyboards}, left. The virtual keyboard was connected through USB with the Android Debug Bridge forwarding the touch positions into the VR application. The layout of the virtual keyboard resembled the standard Android keyboard on the Google Nexus 10, but with the right shift and enter keys switched. This was done to prevent accidental completion of an entered text phrase when users actually wanted to delete characters using the backspace key.

\subsection{Calibration Data Collection}
The calibration phase consisted of three parts: text entry profiling, interpupillary distance (IPD) calibration and finger tip calibration. In the text entry profiling phase, participants were asked to copy prompted sentences using two de-facto standard text entry methods: a desktop keyboard and a QWERTZ touchscreen keyboard. 
Stimulus phrases were shown to the participants one at a time and were kept visible throughout the typing task. 
Participants were asked to type them as quickly and as accurately as possible. They typed stimulus phrases for 3 min using a desktop keyboard and, after a short break, 3 min using a QWERTZ touchscreen keyboard. 
The order of the text entry methods was balanced across participants. 
The interpupillary distance (IPD) was determined using the Oculus IPD calibration tool provided with the Oculus Runtime Environment. The IPD was then used for setting the correct camera distance for stereo rendering. 
For finger tracking, individual retroreflective markers were attached at the nails of participants using double-sided adhesive tape, see Figure \ref{fig:conditionsexternal}, right. The finger calibration aimed at determining the offset between the tracked 3D position of each finger tip and its corresponding nail-attached marker. To this end, the participants were asked to hit three soft buttons of decreasing size (large: 54 $\times$ 68 mm, medium: 35 $\times$ 50 mm, small: 15 $\times$ 15 mm) on the Nexus 10 touch surface while in VR. Initially, the virtual finger tips were shown at the registered 3D positions of the retroreflective markers. On touchdown, the virtual finger tips were transformed by the offset between the 3D coordinate of the touch point and the retroreflective marker. The final positions of the virtual finger tips were averaged across three measurements. Then the participants verified that they could actually hit targeted keys using their virtual finger tip. If necessary, the process was repeated. This calibration procedure was conducted for each finger individually. 

\subsection{Procedure}
The order of the conditions was balanced across participants. In either condition, participants were shown a series of stimulus sentences. For an individual stimulus sentence, participants were asked to type it as quickly and as accurately as possible. Participants were allowed to use the backspace key to correct errors.  
Participants typed stimulus sentences for 15 minutes each condition. The conditions were separated by a 5-minute break, in which participants filled out a the SSQ simulator questionnaire \cite{kennedy1993simulator}, the NASA TLX questionnaire \cite{hart1988development}, and the IPQ \cite{regenbrecht2002real} spatial presence questionnaire.

The system visualised participants' finger tips as semi-transparent spheres. 
By virtue of being a desktop keyboard, participants also benefited from tactile sensation feedback of the keys which also allowed them to perceive when they had typed an individual key.

In the \textsc{TouchscreenKeyboard} condition, when the user touched a key, the key would light up in a green color. The user could at this point prevent the key from being registered as inputted by the system by sliding the finger away from the key and any other keys on the keyboard and then lifting up the finger. If the user instead remained on the key when lifting up the finger, the key was registered as inputted by the system.

\subsection{Results}
Unless otherwise specified, statistical significance tests were carried out using general linear model repeated measures analysis of variance with Holm-Bonferroni adjustments for multiple comparisons at an initial significance level $\alpha = 0.05$. We indicate effect sizes whenever feasible ($\eta^2_p$).

\subsubsection{Entry Rate}
\begin{figure}
	\centering
	\includegraphics[width=\columnwidth]{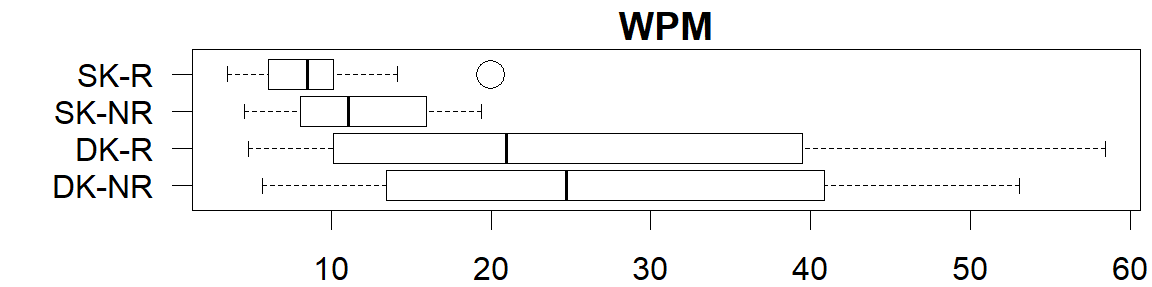}
	\includegraphics[width=\columnwidth]{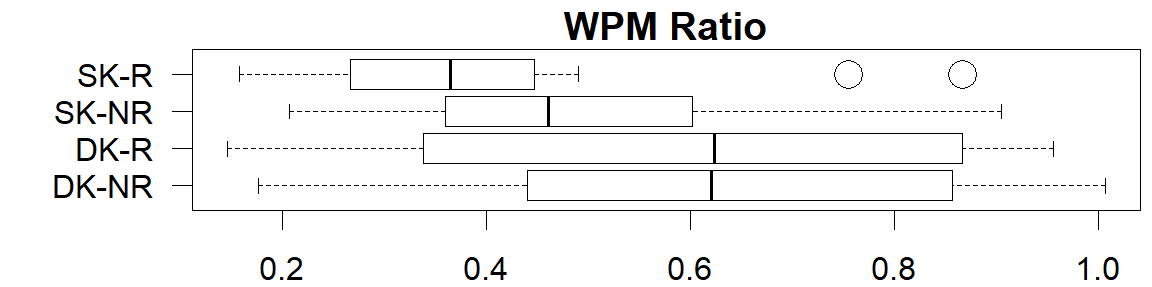}
	\includegraphics[width=\columnwidth]{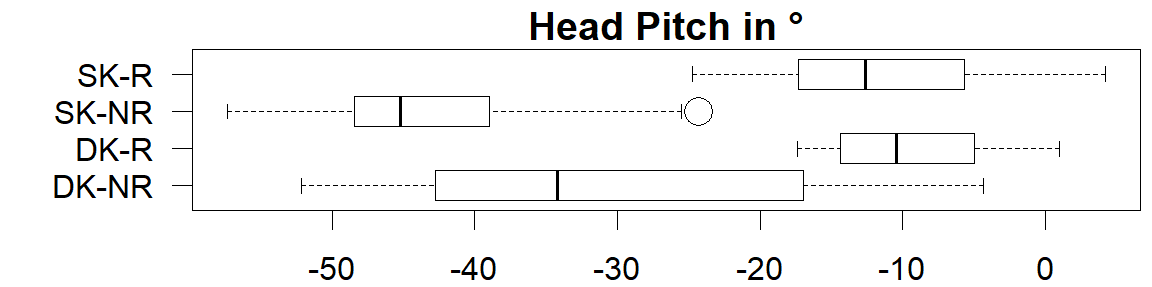}
	\caption{Top: WPM of VR conditions. Middle: WPM ratio between main VR writing phases and non-VR profiling phase. Bottom: Head pitch in degrees, where 0\textdegree~is looking straight ahead and -90\textdegree~is looking straight down. DK-NR: \textsc{DesktopKeyboard+NoReposition}, DK-R: \textsc{DesktopKeyboard+Reposition}, SK-NR: \textsc{TouchscreenKeyboard+NoReposition}, SK-R: \textsc{TouchscreenKeyboard+Reposition}.}	
	\label{fig:wpmratio}
	\vspace{-0.3cm}
\end{figure}

Entry rate was measured in words-per-minute (wpm), with a word defined as five consecutive characters, including spaces. 
For \textsc{DesktopKeyboard}, the mean entry rate was 26.3 wpm (sd = 15.7) in the \textsc{NoReposition} condition and 25.5 wpm (sd = 17.3) in \textsc{Reposition}. For \textsc{TouchscreenKeyboard}, the mean entry rate was 11.6 wpm (sd = 4.5) in the \textsc{NoReposition} and 8.8 wpm (sd = 3.7) in \textsc{Reposition}, see Figure \ref{fig:wpmratio}, top.

The reduction in entry rate from the non-VR profiling phase compared to the VR conditions is visible in Figure \ref{fig:wpmratio}, middle, as the ratio between entry rate in the individual conditions and entry rate in the profiling phase. On average, \textsc{DesktopKeyboard+NoReposition} resulted in a 59\% entry rate compared to profiling, \textsc{DesktopKeyboard+Reposition} in 57\% , \textsc{TouchscreenKeyboard+NoReposition} in 49\% and \textsc{TouchscreenKeyboard+Reposition} in 47\%, with an average baseline text entry performance in the profiling phase of 41.4 word per minute (sd = 12.24) for the desktop keyboard and 23.98 words per minute (sd = 4.17) for the touchscreen keyboard.

The entry rate difference between \textsc{DesktopKeyboard} and \textsc{TouchscreenKeyboard} was, as expected, statistically significant ($F_{1,23} = 25.480$, $\eta^2_p = 0.526$, $p < 0.0001$). However, the interaction between \textsc{KeyboardType} and \textsc{TextPosition} was not significant ($F_{1,23} = 1.936$, $\eta^2_p = 0.078$, $p = 0.177$).
For \textsc{TouchscreenKeyboard}, the difference in entry rate between \textsc{NoReposition} and \textsc{Reposition} was statistically significant ($F_{1,23} = 10.906$, $\eta^2_p = 0.322$, $p = 0.03$). Participants typed significantly faster in the \textsc{NoReposition} condition with an average improvement of 2.8 wpm (32\% relative).
For \textsc{DesktopKeyboard}, the difference in entry rate between \textsc{NoReposition} and \textsc{Reposition} was not statistically significant ($F_{1,23}=0.402$, $\eta^2_p=0.017$, $p=0.532$).  There was no significant improvement in learning.

As proposed by McGill et al. \cite{mcgill2015dose}, we also looked at the time to first keypress. The mean time to first keypress was 1.28  seconds (sd = 0.52) for \textsc{DesktopKeyboard+NoReposition},  1.46 seconds (sd = 0.73) for \textsc{DesktopKeyboard+Reposition},  2.12 seconds (sd = 0.69 ) for \textsc{TouchscreenKeyboard+NoReposition} and 2.61 seconds (sd = 0.96) for \textsc{TouchscreenKeyboard+Reposition}. A repeated measures analysis of variance on the log-transform durations revealed that there was no significant difference ($F_{3,69}=1.951$,$\eta^2_p=0.078$,$p=0.130$).

\subsubsection{Error Rate}
Error rate was measured as character error rate (CER). 
CER is the minimum number of character-level insertion, deletion and substitution operations required to transform the response text into the stimulus text, divided by the number of characters in the stimulus text. 

For \textsc{DesktopKeyboard}, the mean CER was 2.1\% in the \textsc{NoReposition} condition and 2.4\% in \textsc{Reposition}. For \textsc{TouchscreenKeyboard}, the mean CER was 2.7\% in the \textsc{NoReposition} and 3.6\% in \textsc{Reposition}. Typically, a CER less than 5\% is acceptable for general typing (specific threshold depends on use case, see e.g.~\cite{lalomia1994user}). The CER difference between \textsc{DesktopKeyboard} and \textsc{TouchscreenKeyboard} was not statistically significant ($F_{1,23} = 2.545$, $\eta^2_p = 0.1$, $p < 0.124$) and the interaction between \textsc{KeyboardType} and \textsc{TextPosition} was not significant ($F_{1,23} = 0.228$, $\eta^2_p = 0.01$, $p = 0.637$). For \textsc{TouchscreenKeyboard}, the difference in CER between \textsc{NoReposition} and \textsc{Reposition} was not statistically significant ($F_{1,23}=1.078$, $\eta^2_p=0.045$, $p=0.310$). For \textsc{DesktopKeyboard}, the difference in CER between \textsc{NoReposition} and \textsc{Reposition} was not statistically significant ($F_{1,23}=0.104$, $\eta^2_p=0.004$, $p=0.750$). The mean baseline character error rates in the profiling phase where 0.9\% for the desktop keyboard and 2.1\% for the touchscreen keyboard. There was no significant improvement in learning.

\subsubsection{NASA-TLX, Simulator Sickness and Spatial Presence}


The median overall TLX rating was 55.83  for \textsc{DesktopKeyboard+NoReposition}, 53.33 for \textsc{DesktopKeyboard+Reposition}, 60.00 for \textsc{TouchscreenKeyboard+NoReposition} and 61.67 for \textsc{TouchscreenKeyboard+Reposition}.  A Friedman's test revealed no significant differences between \textsc{NoReposition} or \textsc{Reposition} for either \textsc{TouchscreenKeyboard} ($\chi^2(1) = 0.0$, $p=1.0$) or \textsc{DesktopKeyboard} ($\chi^2(1) = 2.667$, $p=0.102$). There were also no significant differences in the sub-scales mental demand, physical demand, temporal demand, performance, effort, frustration.



The median nausea scores were 1.5 for \textsc{DesktopKeyboard+NoReposition} (oculo-motor: 5),  2 for \textsc{DesktopKeyboard+Reposition} (oculo-motor: 4.5), 3 for \textsc{TouchscreenKeyboard+NoReposition} (oculo-motor: 6) and 3 for \textsc{TouchscreenKeyboard+Reposition} oculo-motor: 7). Friedman's test revealed no significant differences.


For spatial presence, the \textsc{TouchscreenKeyboard}  median scores on a 7-item Likert scale were 3.6 
for \textsc{NoReposition} and 3.5 
for \textsc{Reposition}. The difference was not statistically significant (Friedman's test; $\chi^2(1) = 0.167$, $p=0.683$). The \textsc{DesktopKeyboard} median scores were 3.2 
for \textsc{NoReposition} and 3.4 
for \textsc{Reposition}. The difference was not statistically significant (Friedman's test; $\chi^2(1) = 0.391$, $p=0.532$).


\subsubsection{Head Motions}
Figure \ref{fig:wpmratio}, bottom shows the differences in head pitch between the conditions. The mean pitch angle for  \textsc{DesktopKeyboard+NoReposition} was -31\textdegree~(sd = 13), for \textsc{DesktopKeyboard+Reposition} -10\textdegree~(sd = 5), for \textsc{TouchscreenKeyboard+NoReposition} -43\textdegree~(sd = 9) and for \textsc{TouchscreenKeyboard+Reposition} -12\textdegree~(sd = 7). The mean differences between \textsc{DesktopKeyboard+NoReposition} and \textsc{DesktopKeyboard+Reposition} was 20\textdegree, between \textsc{TouchscreenKeyboard+NoReposition} and \textsc{TouchscreenKeyboard+Reposition} 30\textdegree. There were statistically significant differences ($F_{3,69}=90.804$,$\eta^2_p=0.798$,$p<0.0001$). Post-hoc analyses revealed all pairwise differences were statistically significant ($p<0.05$) except between \textsc{DesktopKeyboard+Reposition} and \textsc{TouchscreenKeyboard+Reposition}. The difference between \textsc{DesktopKeyboardNoReposition} and \textsc{TouchscreenKeyboardNoReposition} can be explained by the fact that participants had less haptic feedback with the touch screen, and, hence, needed to look towards the touchscreen keyboard more often.


\subsubsection{Preferences and Open Comments}
Participants were asked to rank the conditions from most preferred (1) to least preferred (4). The median preference rating on a scale from 1 (best) to 4 (worst) was 2 for  \textsc{DesktopKeyboard+NoReposition} as well as for \textsc{DesktopKeyboard+Reposition}, 3 for \textsc{TouchscreenKeyboard+NoReposition} as well as for \textsc{TouchscreenKeyboard+Reposition}. The difference was statistically significant (Friedman's test; $\chi^2(3) = 23.150$, $p<0.001$). Post-hoc analysis with Wilcoxon signed-rank tests and Holm-Bonferroni correction revealed that the general preference for \textsc{DesktopKeyboard} in favour of \textsc{TouchscreenKeyboard} was significant ($p < 0.0001$). All other pairwise differences were insignificant.

Ten participants explicitly mentioned that they preferred \textsc{Reposition} over \textsc{NoReposition} (independent of \textsc{DesktopKeyboard} or \textsc{TouchscreenKeyboard}). Of those, six mentioned that the head position was more comfortable, three mentioned that the text readability was better and one mentioned that with the repositioned keyboard he felt more immersed in VR. Twelve participants mentioned to prefer \textsc{NoReposition} over \textsc{Reposition}. Six participants preferred \textsc{NoReposition} due to habit in normally writing in this configuration, four found it easier to write in this mode and two indicated that this mode was less strenuous. Two participants were indifferent regarding the preference between \textsc{Reposition} and \textsc{NoReposition}.


\section{Discussion}

In our experiment, we saw that users' ability to type on standard keyboards is transferred to VR with about \textasciitilde50\% performance loss. Please note, that this experiment was executed without any automated error-correction, and allowed users to backspace and correct their input at their leisure.
Recent works showed that use of automatic auto-correction may speed up text entry to speeds equivalent to real world typing. Yet, for our experiment, we avoided dependency on one. Beside a simplification of our experiment system design, with no need to implement a decoder, train an appropriate statistical language model, tune model parameters, and provide literal fallback support for text with high perplexity\footnote{Perplexity is defined as $2^H$, where $H$ is entropy.}, such as user names and passwords, text entry in VR may be very context dependent (for example, imagine typing within a context of a Sci-Fi game), where auto correction sufficiency may be limited.

The desktop keyboard text entry rate is about twice faster as the fastest text entry in the literature without auto-correction \cite{Bowman:2004:UIT}, and touchscreen keyboards are equivalent to entry of text using a tracked stylus; a reasonable result given that most users entered text using a single finger. Even modest entry rates of 10-30 wpm are faster than the current text entry existing on commercial VR systems, and may be adequate for many tasks. As a reference point, typing an email consisting of 50 words with an average word length of five characters, including spaces, takes 5 minutes at an entry rate of 10 wpm and 1.7 minutes at an entry rate of 30 wpm. Furthermore, this is done with little or no user learning, using common available hardware, thus removing a potential acceptance barrier. 

\subsection{Touchscreen vs. Desktop Keyboard}

Our results confirm that touchscreen keyboards are significantly slower than desktop keyboards. There are, however, design trade-offs between touchscreen keyboard and desktop keyboard that need to be made explicit in the design process.
At least three design dimensions need to be considered: 1) typing speed, 2) versatility and 3) form factor. 
The desktop keyboard is faster than a touchscreen keyboard. In contrast, a desktop keyboard is more limited than a touchscreen tablet, which can be provided to users in a variety of sizes and shapes. In addition, a touchscreen tablet is more versatile as its user interface can be easily reconfigured to suit different contexts, for instance, in a game, the touch tablet user interface can reveal different buttons, sliders and other user interface widgets depending on the task in the game. As our results indicate, such direct control in VR via a touch tablet becomes feasible when users are provided visual feedback of their finger positions. Alternatively, the user interface, including the keyboard, can adapt to individual users \cite{himberg2003line,gajos2004supple}.


\subsection{Reposition of Hands and Keyboard}


Repositioning the rendering of the keyboard and user's hands from their physical location toward the user view direction can have several benefits, but also comes at potential costs:
First, repositioning could allow users to type in context of the VR world. For example, typing content taken from the environment could be possible without a need to rotate the view away
. While the keyboard and hand representation might block other parts of the VR scene, the position of the keyboard could be chosen such that it would minimize any occluding of important information in the virtual set. In contrast, leaving the position of the desktop keyboard and hands unchanged can result in temporarily blocking of the scene for non-touch typists as they look down at the keyboard representation.  
Still, the virtual environment in our experiment, induced no need to interact with other objects than the keyboard, as we wanted to concentrate on analysis of typing performance in a single task scenario. Hence, future work should investigate if these potential benefits actually manifest themselves in various VR scenes. 

Second, repositioning can have the potential to enable better ergonomics  (20\textdegree~for difference for  \textsc{DesktopKeyboard} and 30\textdegree~for \textsc{TouchscreenKeyboard} conditions) as is shows the keyboard and the user's hands in the user's view direction, without a need for uneasy tilt down of the user's head. 
This notion is also supported by qualitative feedback of our participants. However, a significant difference in term of subjective feedback using Nasa TLX could not be indicated. This could either hint at a negligible effect of \textsc{VirtualKeyboardPosition} on strain, or that the repositioning induces other sources of strain, e.g., a higher coordination effort between the human's visual and motor systems or lower feeling of embodiment in the VR scene.


Finally, displaying the hand and keyboard near the center of the display, and oriented toward the user, enhances the visible quality of the text and keys
. While new VR HMDs might improve the visual quality in the corner areas compared to our system, optical performance will stay better in the center of an HMD lens system for a foreseeable time.

Our results show that for the desktop keyboard, the cost of repositioning the visual feedback of the desktop keyboard and the users' fingers are not significant, while touchscreen keyboards show significant text entry performance degradation. One possible reason for this may be the requirement of touchscreen keyboard for the user to disconnect her fingers from the touch surface to signal key selection. Repositioning the keyboard include a rotation toward the user, which also rotates the visible motion of the fingers, which in turn might slow the user actions. However, we do believe, given the current achieved rate and the merits of repositioning, it might be beneficial to users of touchscreen keyboards too.

\subsection{Limitations and Future Work}


Our study focused on specific items in a large design space of how to design text entry systems in VR. For our evaluations, we focused on the  scenario of a user sitting in front of a desk doing extensive text entry. One reason was to measure text input rate at its limit. Another reason was the observation that this configuration is still popular by many VR applications that do not require the user to walk.  In particular, we can see a great potential of VR as a continuous unlimited VR display, replacing all physical screens in an office environment, supporting both 2D and 3D applications and visualizations. In this scenario, there is need for a robust text entry, which we believe can be filled by current keyboards with additional sensors for hand rendering.

Alternatively, there are many mobile scenarios which could benefit from efficient text entry techniques, also for shorter text sequences. Here, either a handheld or arm-mounted touch screen might serve as a suitable interaction device. In this context, future work should investigate recent mobile text entry techniques for VR, e.g. based on gesturing  \cite{yeo2017investigating}. 

To support our study aims, our virtual environment was designed to have minimal distractions. Future work could investigate how typing is influenced by more engaging virtual scenes and in dual task scenarios (e.g. spatial manipulation of objects in accordance with typing). For example, the effect of repositioning on embodiment or the effect of hand representation on immersion could be studied in further VR scenes. In this context, touch screens might serve this dual purpose better than a desktop keyboard, as, potentially, scene manipulations and typing could be achieved without the need to switch input devices. 

Also, we relied on high precision stationary optical tracking system. But even with this system, we did not sense the movement of physical key presses. The display of the fingers as as they move while typing may help people that do not touch type. The use of mobile depth sensors for hand tracking such as the Leap Motion could be a viable alternative, but their input accuracy for typing would need to be studied. Besides analyzing head movements, future work could also investigate gaze patterns to study if less eye rotations occur in a repositioned keyboard visualization.

\section{Conclusions}

We have examined the use of the standard desktop keyboard and touchscreen keyboard as text entry devices for HMD-based VR applications. We have shown that with a simple rendering of the users' finger tips we can transfer about 50\% of users' typing performance for desktop keyboard to VR, and maintain a comparable performance using touchscreen keyboards, without the need for substantial user learning. Also, we have shown that one can reposition the keyboards and the user's hands, displaying them in front of the user view direction and maintain reasonable performance. Such an ability opens up new opportunities: for example, the standard keyboard can be used, and the user's hands can be rendered at a position of a keyboard or a keypad in the scene in the intended context. 

We believe that VR may expand from the current use of immersive experiences, to a work tool even in the common office, allowing information workers to interact and observe data without the limitation of physical screens. One barrier for such a vision is a robust text entry and editing tool, and we hope this work will be a step in this direction.

\section*{Acknowledgments}
Per Ola Kristensson was supported by EPSRC (grant number EP/N010558/1). We thank the volunteers in the experiment and the anonymous reviewers for their feedback.

\balance{}

\bibliographystyle{abbrv}
\bibliography{vrte}
\end{document}